\documentclass{raa}

\usepackage{graphicx,times}             
\usepackage{natbib}
\usepackage{amssymb,amsmath}
\bibpunct{(}{)}{;}{a}{}{,}

\usepackage[pagebackref=true]{hyperref}

\newcommand{\source}{J0722}

\begin{document}

  \title{TCP\,J07222683$+$6220548: a new AM\,CVn type system with infrequent outbursts}

   \volnopage{Vol.0 (20xx) No.0, 000--000}      
   \setcounter{page}{1}          

   \author{Alexander Tarasenkov\inst{1, 2}
          \thanks{E-mail: tarasenkov@inasan.ru}
          \and
          Kirill Sokolovsky\inst{3}
          \thanks{E-mail: kirx@kirx.net}
          \and
          Alexandr Dodin\inst{2}
          \and
          Oxana Chernyshenko\inst{4}
          \and
          Stanislav Korotkiy\inst{4,5}
          \and
          Ivan Strakhov\inst{2}
          \and
          Marina Burlak\inst{2}
          \and
          Sergey Naroenkov\inst{1}
          \and
          Franz-Josef Hambsch\inst{6,7,8}
          \and
          Tam\'as Tordai\inst{9}
          \and
          Hiroshi Itoh\inst{10}
          \and
          Yasuo Sano\inst{11,12,13}
          \and
          Yusuke Tampo\inst{14,15}
          \and
          Ferdinand\inst{3}
          }

   \institute{
Institute of Astronomy of the Russian Academy of Sciences, 48 Pyatnitskaya Str., Moscow 119017, Russia\\
              \and
Sternberg Astronomical Institute, Lomonosov Moscow State University, Universitetsky Pr., 13, Moscow 119234, Russia\\
              \and
Department of Astronomy, University of Illinois at Urbana-Champaign, 1002 W. Green Street, Urbana, IL 61801, USA\\
              \and
Astroverty, Nizhny Arkhyz, Karachay-Cherkessia, Russia
              \and
Ka-Dar, Nizhny Arkhyz, Karachay-Cherkessia, Russia
              \and
Vereniging Voor Sterrenkunde (VVS), Zeeweg 96, 8200 Brugge, Belgium\\
              \and
Groupe Europ\'{e}en d'Observations Stellaires (GEOS), 23 Parc de Levesville, 28300 Bailleau l'Ev\^{e}que, France\\
              \and
Bundesdeutsche Arbeitsgemeinschaft f\"{u}r Ver\"{a}nderliche Sterne (BAV), Munsterdamm 90, 12169 Berlin, Germany\\
              \and
AAVSO Observer\\
              \and
Variable Star Observers League in Japan (VSOLJ), 1001-105 Nishiterakata, Hachioji, Tokyo, 192-0153, Japan\\
              \and
Variable Star Observers League in Japan (VSOLJ), Nishi juni-jou minami 3-1-5, Nayoro, Hokkaido, Japan\\
              \and
Observation and Data Center for Cosmosciences, Faculty of Science, Hokkaido University, Kita-ku, Sapporo, Hokkaido 060-0810, Japan\\
              \and
Nayoro Observatory, 157-1 Nisshin, Nayoro, Hokkaido 096-0066, Japan\\
              \and
South African Astronomical Observatory, PO Box 9, Observatory, 7935, Cape Town, South Africa\\
              \and
Department of Astronomy, University of Cape Town, Private Bag X3, Rondebosch 7701, South Africa\\
\vs\no
   {\small Received 202x month day; accepted 202x month day}}

\abstract{We present the discovery of TCP\,J07222683$+$6220548, a new ultracompact binary system of the AM\,CVn type. 
This system was first identified displaying a $\Delta V = 7.6$\,mag outburst on 2025-01-20.9416~UTC by the New Milky Way 
wide-field survey for transients and later independently detected by ASAS-SN and ZTF. 
The outburst peaked at $V_{\rm max} = 12.45$ and lasted for seven days, 
followed by a series of rebrightenings. No previous outbursts are found in archival data. 
Positive superhumps with a period of 
 $0.032546 \pm 0.000084$\,d ($46.87 \pm 0.12$\,min), 
barely detectable during the main outburst, became clearly visible during the 
first rebrightening that lasted from day 18 to day 24 after the initial outburst. 
No convincing change in the superhump period was detected. 
Dense time-series photometry follow-up by a pair of 0.5-m INASAN robotic telescopes, 
together with VSNET and AAVSO observers, was essential for identifying TCP\,J07222683$+$6220548 
as an AM\,CVn system and triggering confirmation spectroscopy with the 2.5-m CMO SAI telescope. 
Some outbursting AM\,CVn systems lacking such detailed follow-up may remain unrecognized 
among the newly discovered cataclysmic variable candidates. 
}

   \authorrunning{A. N. Tarasenkov, K. V. Sokolovsky, A. V. Dodin et al. }            
   \titlerunning{TCP\,J07222683$+$6220548: new AM\,CVn system}  

   \maketitle

\section{Introduction}
\label{sec:intro}

The AM Canum Venaticorum (AM\,CVn) stars are a rare class of ultracompact binary systems consisting of 
a white dwarf accreting helium-rich material from a degenerate or semi-degenerate companion 
\citep{1995Ap&SS.225..249W,2010PASP..122.1133S}. 
The nature of this companion (the donor star) can be either a low-mass white dwarf 
(fully degenerate supported by electron degeneracy pressure), or a helium-rich star that is only partially degenerate. 
In the latter case, the star's core is degenerate but its outer layers still experience some thermal pressure support.
These remarkable binary systems have extremely short orbital periods ranging from
$\sim5$ to $\sim70$ minutes, significantly below the $\sim80$ minute period minimum 
of hydrogen-rich cataclysmic variables \citep[e.g.,][]{1985SvAL...11...52T,2009MNRAS.397.2170G}. 
The spectra of AM\,CVn stars are characterized by the presence of helium lines and a notable absence of hydrogen, 
reflecting the evolved nature of the donor star.

Currently, about 70 confirmed and candidate AM\,CVn systems are known
\citep{2022A&A...668A..80L}, with the expectation that there is a large
undiscovered population \citep{2018A&A...620A.141R}. 
The AM\,CVn population can be divided into subgroups based on their orbital periods and accretion behavior \citep{2006ApJ...640..466B}. 
Systems with periods below $\sim 10$ minutes likely undergo direct impact accretion without forming a disk. 
Those with periods between $\sim10$ to 20 minutes maintain stable hot accretion disks and show steady emission. 
The intermediate period systems ($\sim$20 to $\gtrsim$45 minutes) experience dwarf nova-like outbursts where their brightness 
increases by 3-5 magnitudes for days to weeks. 
%
Like hydrogen-rich dwarf novae, these outbursts occur when the accretion disk switches between faint (low-viscosity, low-accretion-rate) 
and bright (high-viscosity, high-accretion-rate) states \citep{2012A&A...545A.115K}, but the instability arises from helium rather than hydrogen ionization 
\citep{1983AcA....33..333S,2012A&A...544A..13K,2020AdSpR..66.1004H,2024A&A...689A.354J}.
The longest period systems 
typically have cool, 
stable disks, although some exceptions are known. 

The formation channels and evolution of AM\,CVn binaries remain subjects of active research
\citep{2015ApJ...805L...6S,2018MNRAS.476.1663G,2019PhDT.......145G,2021ApJ...910...22L,2023MNRAS.520.3187S,2023A&A...678A..34B}. 
They may form through three possible routes: from detached double white dwarf binaries that survive the onset of mass transfer, 
from binaries with helium star donors, or from evolved cataclysmic variables that have lost their hydrogen.
Accumulation of the accreted helium-rich material on the surface of the
primary white dwarf should eventually lead to a helium nova
\citep{1989ApJ...340..509K,1991ApJ...370..615I,2004A&A...419..645Y}.
AM\,CVn stars are also potential progenitors of normal Type Ia and sub-luminous .Ia supernovae
\citep{2007ApJ...662L..95B,2010PASP..122.1133S,2024MNRAS.527.2072D}. AM\,CVn systems may be
related to R Coronae Borealis stars as their progenitors \citep{1996ASPC...96..309S} 
or by being an alternative outcome of common envelope evolution of a pair of white dwarfs
\citep{1984ApJ...277..355W,1996ASPC...96..309S}.

AM\,CVn binaries are expected to be strong sources of low-frequency gravitational waves 
\citep{1987SvA....31..228L,2022A&A...668A..80L,2022ApJ...935....9C}.
detectable by the future space-based gravitational-wave observatories such as 
Laser Interferometer Space Antenna \citep[LISA;][]{2023LRR....26....2A} 
and
TianQin \citep{2021PTEP.2021eA107M}
missions.
The known AM\,CVn systems will serve as guaranteed verification sources for LISA, 
while the mission may potentially discover many more such systems through their gravitational wave emission
\citep{2018MNRAS.480..302K}.

Increasing the sample of known AM\,CVn stars, particularly those showing outbursts, is crucial for several reasons. 
First, it helps constrain their space density and better understand their formation channels. 
Second, outbursting systems provide opportunities to study the physics of helium accretion disks and the mechanisms driving their outbursts. 
Third, a larger sample improves our ability to characterize the low-frequency gravitational wave background 
that will be important for future space-based gravitational wave detectors.

In this paper, we report the discovery and follow-up observations of a new AM\,CVn system identified through its outburst activity. 
The paper is organized as follows: 
Section~\ref{sec:obs} describes the discovery technique and follow-up observations,
Section~\ref{sec:discussion} puts the newly discovered system in the context
of other AM\,CVn stars and Section~\ref{sec:conclusion} summarizes our findings.


\section{Observations and analysis}
\label{sec:obs}

\subsection{NMW survey discovery}
\label{sec:nmw}

The New Milky Way survey\footnote{\url{https://scan.sai.msu.ru/nmw/}} \citep[NMW;][]{2014ASPC..490..395S} aims to rapidly detect
bright Galactic transients including classical novae, dwarf novae, flare
stars, young stellar object FUOR/EXOR outbursts \citep[for reviews see][]{2014prpl.conf..387A,2016ARA&A..54..135H} and 
brightest microlensing events \citep[e.g.,][]{2020A&A...633A..98W} in anticipation of more exotic transients such as 
a luminous red nova \citep{2011A&A...528A.114T,2014MNRAS.443.1319K,2022MNRAS.517.1884A}, 
a helium nova \citep{2003A&A...409.1007A,2021MNRAS.501.1394N} and ultimately the next Galactic
supernova \citep{2013ApJ...778..164A}. The ambition of the survey is to 
inform the community of the appearance of an astrophysically interesting 
transient over the minimal possible time.

The NMW survey operates two wide-field CCD cameras: unfiltered monochrome ST-8300M and STL-11000M
attached to identical Canon 135\,mm f/2.0 telephoto lenses and installed on computer-controlled HEQ5~Pro mounts. 
The two cameras have the field of view (pixel scale) of 
$7.7\degr{}\times5.8\degr{}$ (8.35\arcsec{}/pix) 
and
$15\degr{}\times10\degr{}$ (13.8\arcsec{}/pix), respectively. 
The cameras are housed in the rolling-roof pavilion at 
the Astroverty astrofarm\footnote{\url{https://astrovert.ru/astrofarm/}} in Nizhnii Arkhyz, Karachay-Cherkessia, Russia. 
The entire sky visible from the observing site is divided into a set of overlapping fields. 
Three 20\,s exposures of each field are taken with dithering to aid in distinguishing celestial objects (fixed relative to 
the stars) from image artifacts (fixed relative to the chip). 
The transient detection pipeline is based on the \textsc{VaST} code \citep{2018A&C....22...28S}. 
It does not utilize the computationally expensive image subtraction and 
instead relies on matching the lists of sources detected on reference and 
second-epoch images. 
%
Apart from the images' limiting magnitude (around 14\,mag on a moonless night), 
detection efficiency is limited by transient blending with surrounding stars. 
Insert-and-recovery simulations demonstrate the system has about 80\% (99\%) chance of detecting a
13\,mag (10\,mag) transient in a low Galactic latitude field. 
After human vetting, the identified new transients are reported via 
the CBAT ``Transient Objects Confirmation Page''\footnote{\url{http://www.cbat.eps.harvard.edu/unconf/tocp.html}} 
with the exception of UV~Ceti type flaring red dwarfs and other
non-cataclysmic variable stars that are reported directly to 
the AAVSO International Variable Star Index \citep[VSX;][]{2006SASS...25...47W}.

The 12.8\,mag transient TCP\,J07222683$+$6220548 
(referred as \source{} in the following) was found in the NMW images obtained on 2025-01-20.9416~UTC
($t_0 = {\rm JD}$\,2460696.4416). 
No objects brighter than $CV=14.5$ 
is visible at the previous NMW image of the field obtained ten days earlier on 2025-01-10.9233. 
The All-Sky Automated Survey for Supernovae \citep[ASAS-SN;][]{2014ApJ...788...48S,2017PASP..129j4502K}
has independently discovered \source{} 36 hours later on 2025-01-22.46 naming it ASASSN-25aj. 
The latest ASAS-SN observation of the field was on 2025-01-20.3957 (13 hours prior to
the NMW detection) showing no object brighter than $g = 17$. 
The transient was also identified as an anomaly among the Zwicky Transient Facility \citep[ZTF;][]{2019PASP..131a8003M}
transients by \citep[][]{2025ATel17058....1L} and designated ZTF25aacfjde.

\subsection{Gaia counterpart, distance and extinction}
\label{sec:posext}

A blue source Gaia\,DR3\,1087501559186713472 (07:22:27.05915 $+$62:20:56.7450
Equinox$=$J2000.0 at Epoch$=$J2000, $BP=20.00 \pm0.06$, $RP=20.01 \pm0.09$,
${\rm Plx}=1.96 \pm0.53$\,mas, ${\rm PM} = 9.9$\,mas/yr; \citealt{2023A&A...674A...1G}) is located 2.5\arcsec{} from the measured position of the transient, 
within the few-arcsecond astrometric uncertainty of the discovery observations
(8.35\arcsec{}/pix image scale). Despite the large pixel scale, 
the identification is unambiguous as this is a relatively uncrowded field. 
The second-closest Gaia~DR3 source is located 36\arcsec{} away from the discovery position.
The second-closest PanSTARRS1 \citep{2016arXiv161205560C} source is 20\arcsec{} away.
The matching source also clearly stands out in both PanSTARRS1 and DSS2 \citep{1996ASPC..101...88L} 
color images as having much bluer color compared to field stars.
The follow-up astrometric measurements reported by K.~Yoshimoto via the TOCP confirm the Gaia source identification.

The Gaia parallax corresponds to the geometric distance of
$575^{+304}_{-153}$\,pc with 68\% uncertainties according to \cite{2021AJ....161..147B}.
Table~\ref{tab:mags} lists the observed and absolute peak and minimum magnitudes of
\source{} together with the extinction corrections. 
The corrections were derived using the \textsc{mwdust} package \citep{2016ApJ...818..130B} that
combines 3D dust maps of \cite{2003A&A...409..205D,2006A&A...453..635M,2019ApJ...887...93G}.
The same extinction values are predicted for the whole 68\% range of distances.
The uncertainty in absolute magnitude is dominated by the uncertainty in distance, so applies equally to all bands.
The minimum $V$ band magnitude, $V_{\rm min}$, was color-transformed from PanSTARRS1
$g$ and $r$ band photometry using the relation from
\citep{2012ApJ...750...99T}:
\begin{equation}
V =  0.474 * (g - r) + 0.006 + r.
\end{equation}
We have used the extinction coefficients from Table~2 of
\cite{2018MNRAS.479L.102C} to compute the extinction in Gaia bands from
$E(B-V)$ provided by \textsc{mwdust}. 

%
\begin{table}
\begin{center}
\caption[]{\source{} Magnitudes and Extinction}
\label{tab:mags}

 \begin{tabular}{cccc}
  \hline\noalign{\smallskip}
Filter &  Observed & Extinction & Absolute \\
  \hline\noalign{\smallskip}
$V_{\rm max}$  & $12.45 \pm 0.01$ & 0.30 &  $3.4^{+0.7}_{-0.9}$ \\ 
$V_{\rm min}$  & $20.06 \pm 0.03$ & 0.30 & 11.0 \\ 
$g$            & $19.98 \pm 0.02$ & 0.35 & 10.8 \\ 
$r$            & $20.12 \pm 0.02$ & 0.25 & 11.1 \\ 
$G$            & $20.01 \pm 0.01$ & 0.27 & 10.9 \\ 
$BP$           & $20.00 \pm 0.06$ & 0.33 & 10.9 \\ 
$RP$           & $20.01 \pm 0.09$ & 0.20 & 11.0 \\ 
  \noalign{\smallskip}\hline
\end{tabular}
\end{center}
\end{table}

\subsection{Spectroscopy with the 2.5-m CMO SAI telescope}
\label{sec:spectroscopy}

\begin{figure*}[t!]
\centering
\includegraphics[width=\textwidth,clip=true,trim=0.0cm 0cm 0.5cm 0.2cm,angle=0]{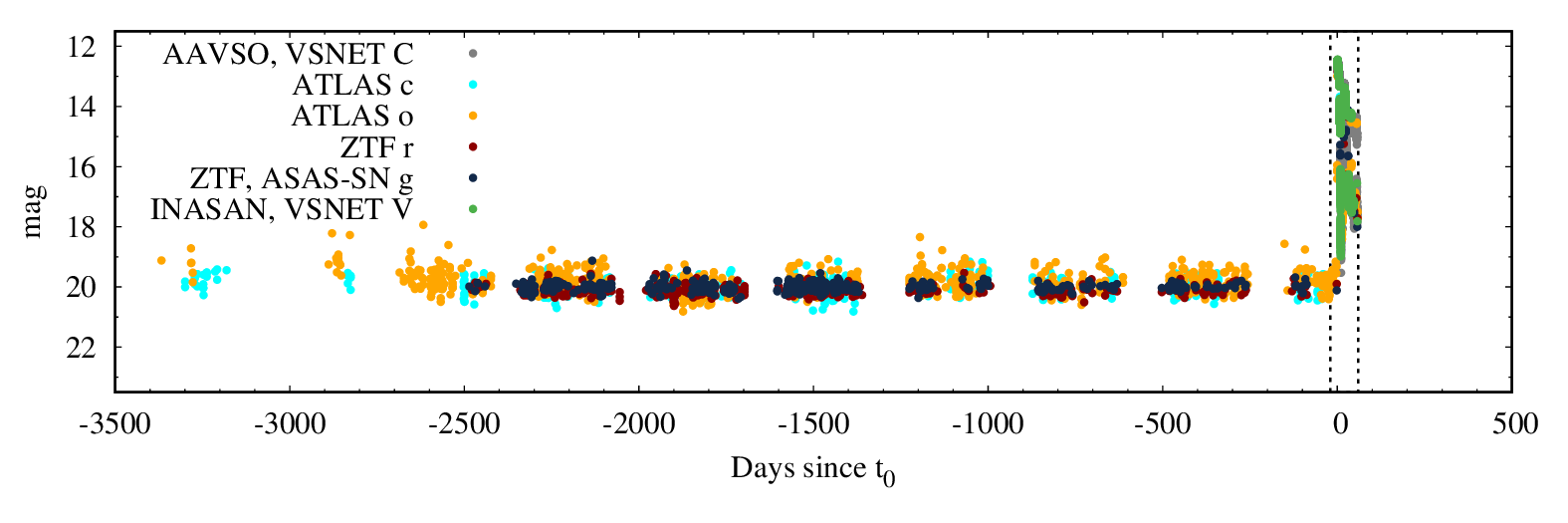}
\includegraphics[width=\textwidth,clip=true,trim=0.0cm 0cm 0.5cm 0.2cm,angle=0]{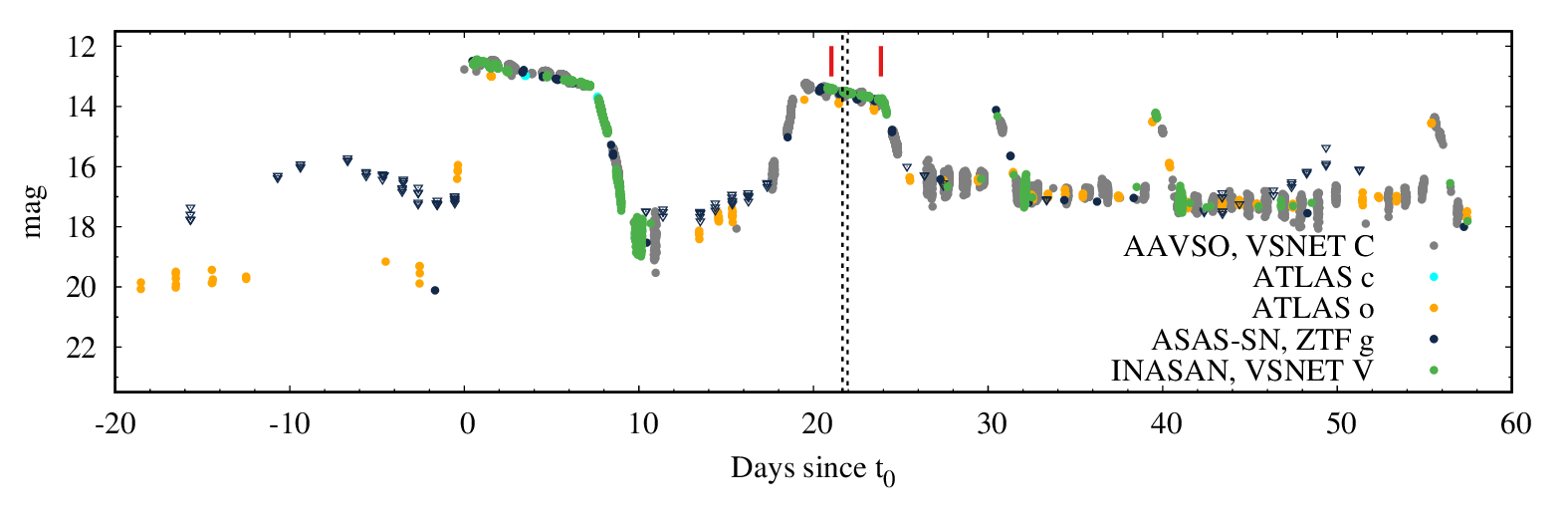}
\includegraphics[width=\textwidth,clip=true,trim=0.0cm 0cm 0.5cm 0.2cm,angle=0]{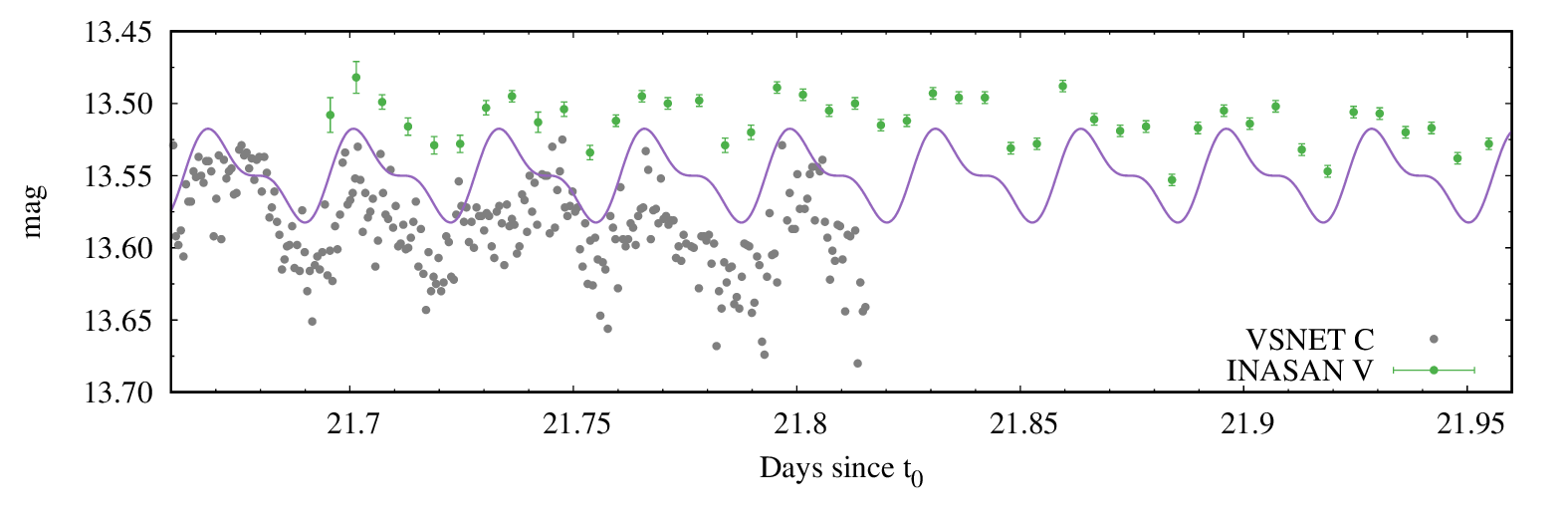}
\caption{The lightcurve of \source{} combining $V$-band photometry from INASAN Terskol and Kislovodsk
telescopes (Section~\ref{sec:photometry}) with $V$-band and unfiltered $CV$-band measurements
collected by VSNET and AAVSO observers (including $CV$ photometry from the two NMW cameras), 
ASAS-SN $g$, ATLAS $o$ (orange) and $c$ (cyan), and ZTF $g$ and $r$ data. 
The top panel shows the complete combined lightcurve with the vertical
dashed lines indicating the outburst plotted in the middle panel.
The vertical dashed lines in the middle panel indicate time range of bottom
plot that presents an example section of the lightcurve where superhumps are
visible (Section~\ref{sec:period}). The purple curve approximates the
superhump shape as a sum of two sine waves, one with a period from
eqn.~(\ref{eq:period}) and the peak-to-middle amplitude of 0.025\,mag and
the other with twice that period and half the amplitude (c.f.~Fig.~\ref{fig:phasedlcbest}), it is plotted to guide the eye.
The outburst plot (middle panel)} includes ASAS-SN $g$ upper limits (triangles) along with the positive detections (filled circles). 
Vertical red bars mark the times of spectroscopic observations described in Section~\ref{sec:spectroscopy}.
\label{fig:lc}
\end{figure*}

Spectroscopic observations of \source{} were performed with 
the 2.5-m telescope at the Caucasian Mountain Observatory (CMO) of 
the Sternberg Astronomical Institute of Lomonosov Moscow State University \citep[SAI MSU;][]{2020gbar.conf..127S}
using the Transient Double-Beam Spectrograph \citep[TDS;][]{2020AstL...46..836P}. 
We used 1\arcsec{} slit width, which provides resolving power of 1300 in the blue channel (3550-5700\,\AA) 
and 2400 in the red channel (5700-7450\,\AA) of the spectrograph.
The spectra were obtained 
on 
2025-02-10.9578 (total exposure time was 600\,sec.)
and 
2025-02-13.8017 (total exposure time was 3600\,sec.) during the rebrightening, Figure~\ref{fig:lc}. 
The spectra were flux calibrated with A0V comparison stars, 
but due to light losses on the narrow slit, the flux may 
be offset 
by a constant factor.

\begin{figure*}
\centering
\includegraphics[width=\textwidth,clip=true,trim=0.0cm 0cm 0.75cm 0.2cm,angle=0]{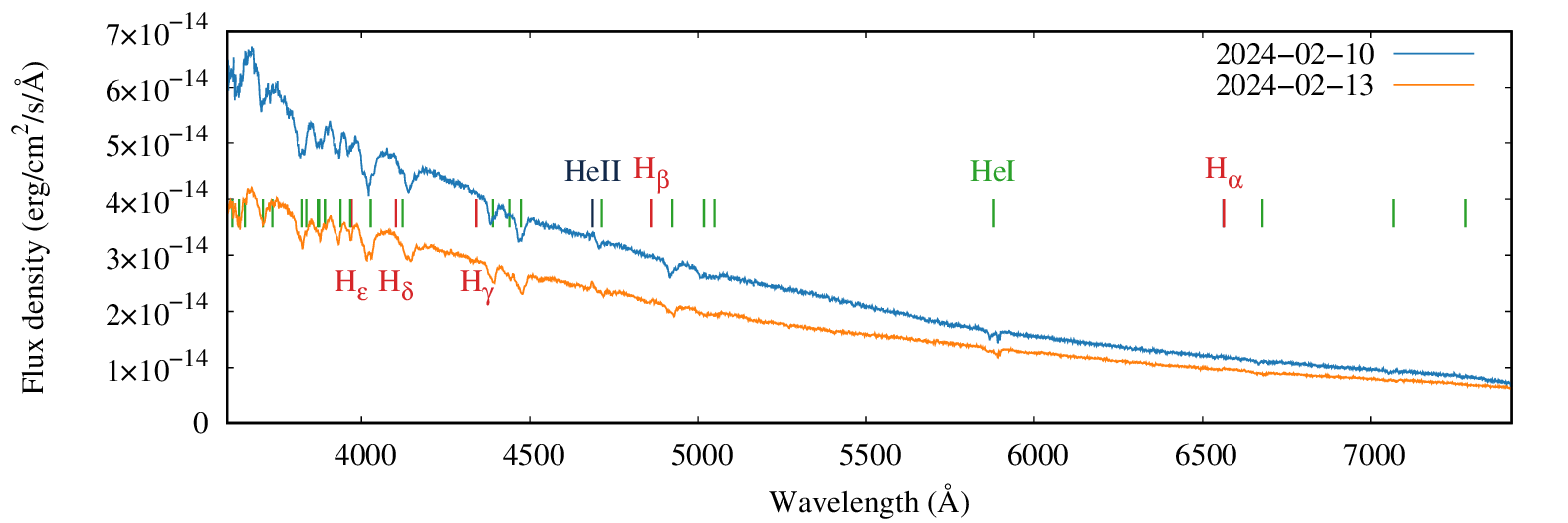}
\caption{The spectra of \source{} obtained with the 2.5-m SAI MSU telescope during the rebrightening.
The green and dark blue marks show the location of prominent He~I and He~II lines, respectively. 
The red marks indicate the expected location of Balmer lines.}
\label{fig:spec}
\end{figure*}

The spectrum (Figure~\ref{fig:spec}) shows a blue continuum with prominent broad absorption lines of He~I. 
Notably, the spectrum lacks the Balmer hydrogen lines. 
The He~II 4687\,\AA{} emission is also present.
The list of He lines was obtained from the NIST Atomic Spectra Database \citep{NIST_ASD}.

Such a spectrum is typical for an AM\,CVn system in outburst \citep[e.g.,][]{2010MNRAS.407.1819R,2013MNRAS.430..996L}. 
Similar to hydrogen-abundant dwarf novae \citep{2010A&A...519A.117I}, AM\,CVn systems show an absorption spectrum
while in outburst and emission line spectrum in quiescence. \cite{2024MNRAS.532.4205P}
in their figure~5 present spectral evolution of AM\,CVn system ASASSN-21br
from outburst to quiescence illustrating the transition from helium
absorption to emission dominated spectrum \citep[see also][]{2007MNRAS.379..176R,2011ApJ...739...68L}.

\subsection{Photometry}
\label{sec:photometry}

Time series photometry of \source{} was performed using multiple small 
telescopes described below. 
The results were shared via the AAVSO International Database \citep{AAVSODATA}, 
the Variable Star Observers League in Japan (VSOLJ) database\footnote{\url{http://vsolj.cetus-net.org/index.html}} 
and the \texttt{vsnet-alert} mailing list. 
These targeted observations were complemented with more sparsely-sampled survey 
data to better trace the overall shape of the lightcurve prior to and during
the outburst.

Photometry of \source{} was performed using two robotic telescopes operated by the Institute of Astronomy of the Russian Academy of Sciences (INASAN).
These instruments are based on Ritchey--Chr\'etien Astrosib RC500 telescopes with the aperture 
0.5-m 
and focal ratio f/8. 
The telescopes are operated using a dedicated control system designed to enable both remote and fully automatic observations \citep{Terskol}. 
The system implements continuous real time monitoring of weather conditions and astronomical alerts - notifications of newly discovered transient events such as gamma-ray bursts (GRBs). 
A centralized observation planning optimizes the distribution of observing tasks between the telescopes.
Observations of \source{} during its main outburst were carried out using a robotic telescope installed 
at Terskol Observatory of INASAN \citep{2024PZ.....44...78T} in the $V$ band. Photometric monitoring was conducted every clear night in automatic mode.
The object's high northern declination allowed continuous observation throughout the night. 
During rebrightening, we performed photometry of \source{} in $BVRI$ bands using
a similar 
0.5-m 
Astrosib RC500 robotic telescope of 
INASAN Kislovodsk Observatory 
\citep{Kislovodsk}. Both telescopes are equipped with a filter wheel with Johnson $UBVRI$ filters. 
The telescope at the Kislovodsk Observatory is equipped with a ZWO ASI6200MM~Pro CMOS camera, 
while the telescope at the Terskol Observatory uses a FLI ProLine~16803 CCD camera.
Aperture photometry was performed using \textsc{AstroImageJ} software \citep{AiJ}. 
The photometric aperture size was adjusted to match changes in seeing during image processing.
Stars from the APASS catalogue \citep{APASS}, which are close to \source{}, were used as comparison.

Photometric observations of \source{} were performed at Andromeda observatory in Mol, Belgium using 
a Carbon tube C14~Edge~HD 0.36-m telescope at f/7 and a QHY600M CMOS camera with Astrodon photometric 
filters. Aperture photometry was performed 
with \texttt{LesvePhotometry} software\footnote{\url{http://www.dppobservatory.net/astroprograms/LesvePhotometryDownloadPage.php}} 
using AAVSO sequence comparison stars. 

\source{} was observed using with a 
0.3-m 
f/4 Newtonian telescope and QHY IC8300 CCD 
remotely operated with \texttt{INDI}\footnote{\url{https://www.indilib.org/}} and \texttt{Ekos}/\texttt{Kstars}. 
Image processing was done by \texttt{gcx}\footnote{\url{https://gcx.sourceforge.net/}} software and aperture photometry was performed by
\texttt{IRAF} \citep{1986SPIE..627..733T} under \texttt{Linux}. 

\source{} was also observed with a 
0.3-m 
Schmidt--Cassegrain telescope
equipped with an unfiltered ZWO ASI294MM Pro CMOS camera. 
The source was also observed using a 
0.2-m 
Schmidt--Cassegrain telescope
with an unfiltered ATIK383L CCD camera. The images from these two telescopes
were measured using \texttt{Muniwin}~2.1.36\footnote{\url{https://c-munipack.sourceforge.net/}}. 

The observations of \source{} were also performed using a 
0.36-m 
Schmidt--Cassegrain telescope
(focal length 3950\,mm, f/11) equipped with a FLI ML1001E CCD camera and a
set of Johnson-Cousins filters. \texttt{MIRA~Pro~64} software was used for photometry.

To better track the long-term evolution of the lightcurve we supplement 
the targeted observations collected with the telescopes described above with 
unfiltered $CV$ band photometry collected by 
the two wide-field NMW cameras (Section~\ref{sec:nmw}). 
The photometry was also extracted from the ASAS-SN Sky Patrol \citep{2014ApJ...788...48S,2017PASP..129j4502K} ($Vg$ bands),
ZTF ($gr$ bands) Public Data Release~23 (covering time interval from March 2018 to June 2023) 
accessed via the \textsc{SNAD Viewer}\footnote{\url{https://ztf.snad.space/}} \citep{2023PASP..135b4503M} 
together with ZTF outburst photometry obtained through the \textsc{Lasair} broker\footnote{\url{https://lasair-ztf.lsst.ac.uk/objects/ZTF25aacfjde/}} \citep{2024RASTI...3..362W} 
and ATLAS \citep[$c$ {\it cyan} and $o$ {\it orange} bands;][]{2018PASP..130f4505T,2020PASP..132h5002S} surveys. 
The ASAS-SN and ATLAS measurements were extracted from original (not difference) images using forced photometry 
at the Gaia~DR3 position of the host system (Section~\ref{sec:posext}). 
The long-term lightcurve of \source{} is presented in Figure~\ref{fig:lc}.

\section{Results and discussion}
\label{sec:discussion}

\subsection{Overall shape of the lightcurve}
\label{sec:lc}


The combined lightcurve of \source{} (Figure~\ref{fig:lc}) displays a single outburst in 2025.
The lightcurve spans 9 years with positive detections at quiescence by ZTF and ATLAS 
and 13 years with ASAS-SN upper limits. The seasonal gaps in the lightcurve 
(smallest for the ASAS-SN data) are typically about 100 days long.
If a previous outburst happened less than 13 years ago, 
in order to be completely missed it would have to start after the start of 
a seasonal gap and end before the end of that gap. 
Taking the outburst duration together with the first rebrightening to be 25\,d 
(see the discussion below and Figure~\ref{fig:lc}), 
the effective gap length becomes 75\,d, corresponding to a 20\% chance of
missing a single outburst and a 4\% chance of missing two outbursts occurring
at random times over the past 13 years. 
If we also consider the post-outburst decline lasting $>60$\,d when the
system was $\sim 2$\,mag brighter than in quiescence (a difference that
should have been easily visible in the ATLAS and ZTF photometry), this further
reduces the probability of completely missing an outburst over the last decade.
We conclude that outbursts in \source{} are likely rare.

The pre-outburst ZTF and ATLAS lightcurve does not show any obvious
variability. Specifically, there seems to be no pre-outburst brightening in
contrast to what was observed in another AM\,CVn system ASASSN-21au by \cite{2022ApJ...926...10R}.
The season-to-season variations hinted by the ATLAS lightcurve of \source{} 
are probably of an instrumental origin as the changes do not match between $c$ 
and $o$ bands and are not seen in ZTF $gr$ photometry.


The 2025 outburst of \source{} is characterized by a fast rise at a rate of 7.8\,mag/d (0.325\,mag/hour), estimated from 
the comparison of ATLAS photometry during the lightcurve rise and the NMW discovery photometry.
The total outburst amplitude is $\Delta V = 7.6$\,mag. 
Such fast rise and high amplitude are typical for dwarf novae \citep[e.g.,][]{2019MNRAS.490.5551R}.
%
After peaking some time between the NMW discovery at $t_0$ and the next NMW
observation 11 hours later, \source{} begins a linear (in magnitudes as a
function of time) decline at a rate of 0.128\,mag/d.
This slow decline continues until $t_0 + 7.2$\,d and by $t_0 + 7.7$\,d
changes to a rapid 
2.1\,mag/d 
decline. The decline stops about 1.25\,mag
above the median quiescence level by $t_0 + 10$\,d.
According to ATLAS photometry, by $t_0 + 13$\,d it changes to a gradual rise of 0.3\,mag/d, 
that around $t_0 + 18$\,d turns into a more rapid rise towards a rebrightening. 
The rise to the rebrightening is happening at a rate 3.5\,mag/d, considerably
slower than the rise toward the main outburst.
%
%
The rebrightening is characterized by a decline rate of 0.123\,mag/d, similar
to that of the main outburst. The end of the rebrightening at $t_0 + 24.1$\,d
is well constrained by the INASAN-Kislovodsk photometry. 
The rapid 
2.0\,mag/d 
post-rebrightening decline stops 3.3\,mag above
quiescence. At least three short-lived rebrightenings follow on 
$t_0 + 30$\,d, $t_0 + 39$\,d and $t_0 + 55$\,d. 
Table~\ref{tab:outburst} summarizes the properties of the main outburst and
the first rebrightening.

\begin{table*}
\begin{center}
\caption[]{Outburst Parameters}
\label{tab:outburst}

 \begin{tabular}{cccccc}
  \hline\noalign{\smallskip}
Outburst          &  Rise Rate & Plateau Rate & Decline Rate & Plateau Duration & Peak \\
                  &  (mag/d)   & (mag/d)      & (mag/d)      & (d)              & (mag) \\
  \hline\noalign{\smallskip}
Main              &  $-7.8$    & 0.128        & 2.1          & 6.7              & 12.45 \\
1st Rebrightening &  $-3.5$    & 0.123        & 2.0          & 4.6              & 13.22 \\
  \noalign{\smallskip}\hline
\end{tabular}
\end{center}
\end{table*}

The magnitude change rates quoted in Table~\ref{tab:outburst} were obtained by robust linear 
fitting \citep[implemented in the \texttt{GNU Scientific Library};][]{gough2009gnu}
to the relevant sections of the color-combined lightcurve.
We do not report fitting uncertainties as the true uncertainties are
dominated by the lightcurve sampling (at what times observations are
available), the accuracy of color correction for the overall lightcurve
(where the same color regardless of magnitude is assumed), and the validity
of linear approximation to the actual lightcurve shape. 
Comparison with weighted linear fits and fits including or excluding individual points
suggest that the magnitude change rates are typically constrained to within
20\% by the available data.

The difference in rise times (rates of brightening) between 
outbursts of an individual dwarf nova are often interpreted as 
the difference between the ``outside-in'' and ``inside-out'' outbursts 
\citep{1986ApJ...301..634C,2001A&A...366..612B,2012A&A...544A..13K,2024A&A...689A.354J}.
The fastest rise is 
attributed to 
an ``outside-in'' outburst: 
the thermal instability begins at the outer edge of the accretion disk and propagates inward
The slower rise is expected for an ``inside-out'' outburst, where the inner part of the accretion disk
is the first to transition to the ionized state, with the
heating/ionization wave propagating to progressively lager disk radii. 
The decay time (rate) is expected to be the same for both ``outside-in'' and ``inside-out''
outbursts, as the cooling wave always moves from the outer region of the disk to the inner 
\citep[section~3.5.4.1 of][]{1995cvs..book.....W}.
Comparing this to the observed rise and decline rates listed in
Table~\ref{tab:outburst}, we suggest that the main outburst might have been
of the ``outside-in'' type, while the first rebrightening was an ``inside-out'' outburst.

The outburst rise and decay rates should also depend on the accretion 
disk size, with longer rise and decay timescales found in systems with longer orbital periods
and hence larger disks \citep[section~3.3.3.5 of][]{1995cvs..book.....W}. 
For \source{}, the first rebrightening, which is characterized by a slower rise, 
is unlikely to be associated with the accretion disk being larger than
during the main outburst, which displayed a faster rise
(Figure~\ref{fig:lc}. Table~\ref{tab:outburst}). 
The fainter peak magnitude of the rebrightening compared to the main outburst suggests that
the disk reached a smaller maximum radius during the rebrightening.

The rebrightenings (sometimes referred to as ``echo outbursts'') and the
elevated (compared to pre-outburst time) brightness level between them may
result from an enhanced mass transfer level from a secondary heated by 
the accreting white dwarf during the outburst \citep{2021A&A...650A.114H}.

An isolated outburst with a week-long plateau bracketed by the rapid rise and decline
phases and followed by rebrightenings is typical for outbursting AM\,CVn
type systems. However, similar lightcurve shapes are found among 
hydrogen-rich dwarf novae of WZ\,Sge type, as discussed in Section~\ref{sec:class}.

\subsection{Periodic modulation}
\label{sec:period}

\begin{figure*}
\centering
\includegraphics[width=0.48\textwidth,clip=true,trim=0.0cm 0.0cm 0.0cm 0.0cm,angle=0]{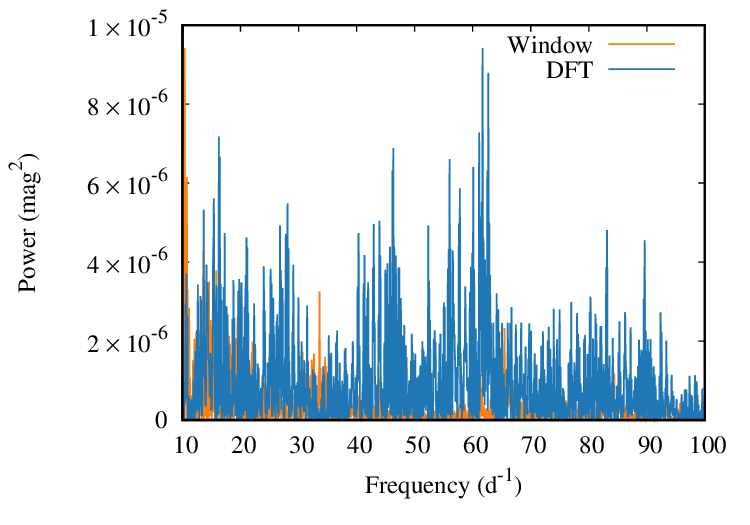}
\includegraphics[width=0.48\textwidth,clip=true,trim=0.0cm 0.0cm 0.0cm 0.0cm,angle=0]{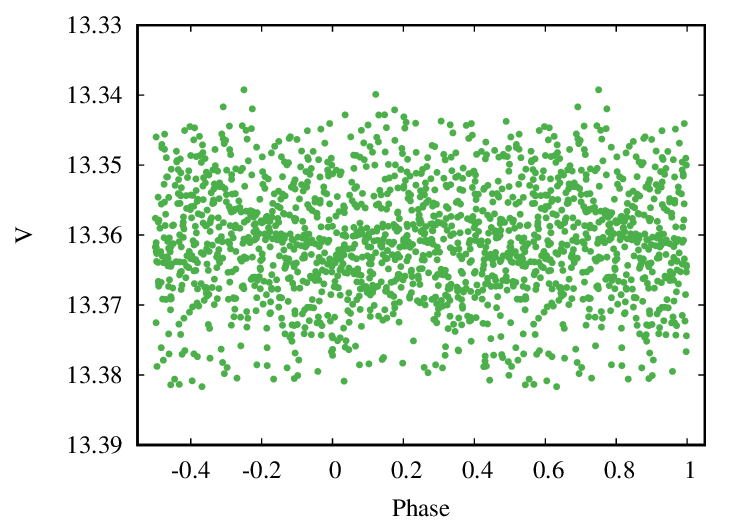}
\includegraphics[width=0.48\textwidth,clip=true,trim=0.0cm 0.0cm 0.0cm 0.0cm,angle=0]{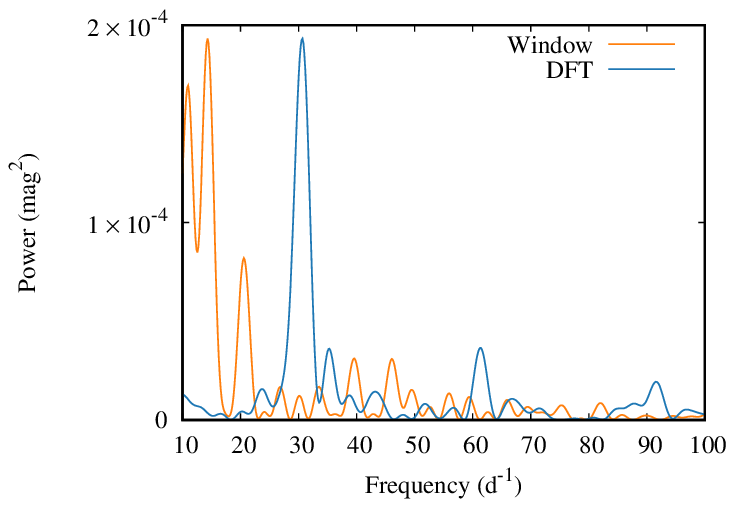}
\includegraphics[width=0.48\textwidth,clip=true,trim=0.0cm 0.0cm 0.0cm 0.0cm,angle=0]{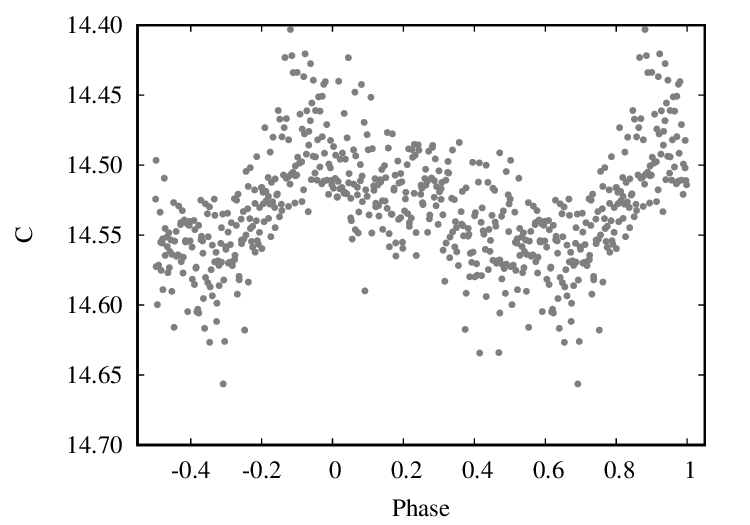}
\includegraphics[width=0.48\textwidth,clip=true,trim=0.0cm 0.0cm 0.0cm 0.0cm,angle=0]{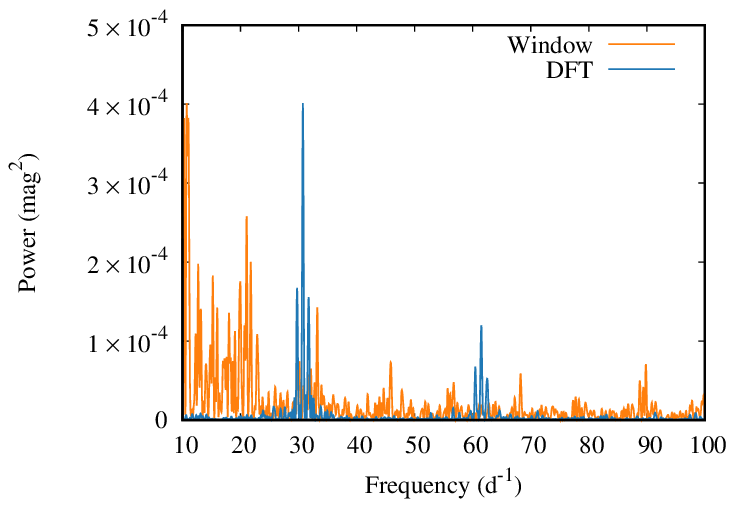}
\includegraphics[width=0.48\textwidth,clip=true,trim=0.0cm 0.0cm 0.0cm 0.0cm,angle=0]{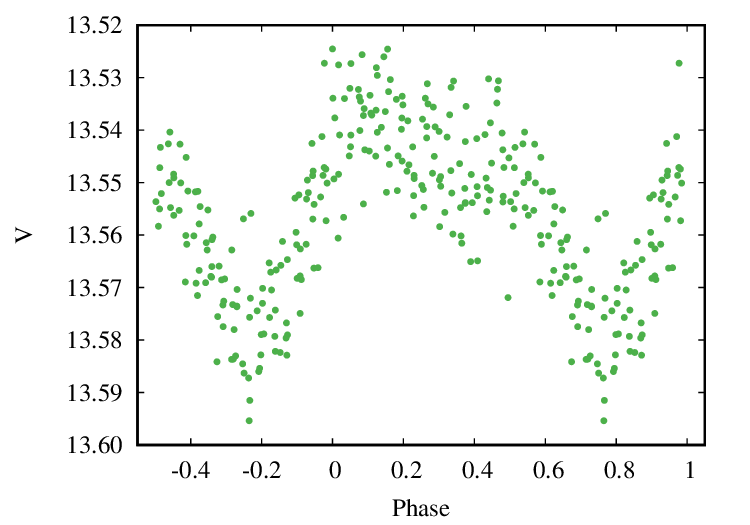}
\caption{Power spectra with their corresponding spectral windows and phased lightcurves of \source{} during 
the main outburst (top panels; INASAN-Terskol; $V$ filter),
the rise to the rebrightening (middle panels; VSNET Hiroshi Itoh; Clear band)
 and during rebrightening (bottom panels). 
All lightcurves are phased with the light elements ${\rm HJD(TT)} = 2460718.3052 + 0.032516 \times {\rm E}$
corresponding to the highest peak in the rebrightening power spectrum
(bottom left panel; INASAN-Kislovodsk; $V$ filter). }
\label{fig:phasedlc}
\end{figure*}

\begin{figure*}
\centering
\includegraphics[width=0.48\textwidth,clip=true,trim=0.0cm 0.0cm 0.0cm 0.0cm,angle=0]{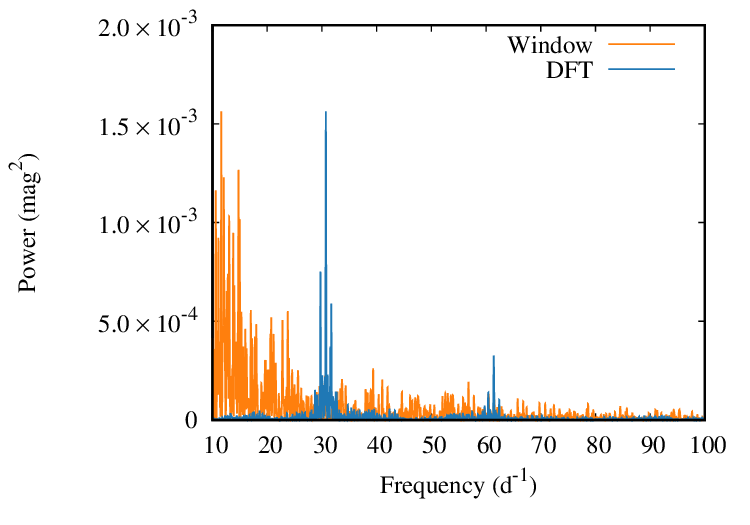}
\includegraphics[width=0.48\textwidth,clip=true,trim=0.0cm 0.0cm 0.0cm 0.0cm,angle=0]{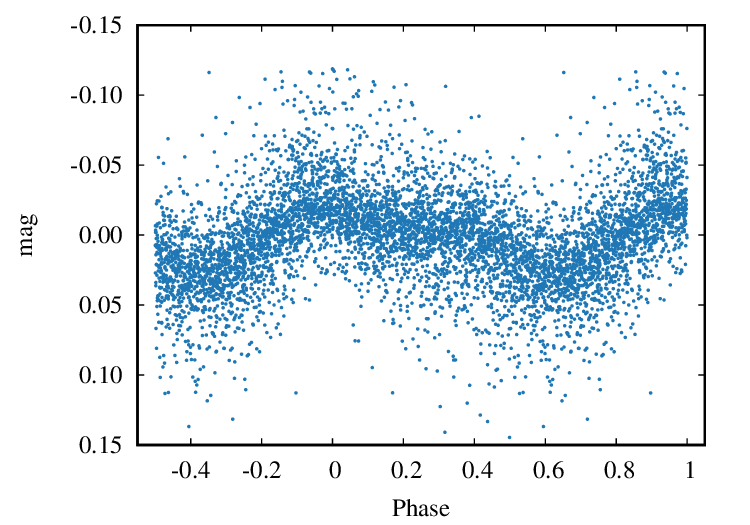}
\caption{Power spectrum with its corresponding spectral window (left panel) and phased lightcurve of \source{}
(right panel) during the rebrightening constructed by combining INASAN-Kislovodsk $BVRI$ and 
VSNET Clear band data. This dataset combines the data presented in 
the middle and lower panels of Figure~\ref{fig:phasedlc} with additional Clear band photometry.
The lightcurve (right panel) is phased with the light elements 
${\rm HJD(TT)} = 2460721.0757 + 0.032546 \times {\rm E}$
corresponding to the highest peak in the rebrightening power spectrum (left panel).}
\label{fig:phasedlcbest}
\end{figure*}

After applying the Heliocentric correction \citep[e.g.,][]{2010PASP..122..935E} 
and detrending the $V$ band lightcurve with a piecewise linear function 
\citep[e.g.,][]{2023arXiv231104903S} we performed a period search using
multiple techniques: 
the Discrete Fourier Transform \citep[DFT;][]{1975Ap&SS..36..137D,2014MNRAS.445..437M,2025A&A...693A.319K}, 
it's Lomb-Scargle modification that allows for analytical computation of false alarm probability
if the photometric measurements are affected by uncorrelated (white) Gaussian noise
\citep{1976Ap&SS..39..447L,1982ApJ...263..835S,2008MNRAS.388.1693F,2018ApJS..236...16V}, 
and the string-length method of \cite{1965ApJS...11..216L} 
implemented in the codes 
\textsc{WinEFK}\footnote{\url{http://www.vgoranskij.net/software/}} \citep{1976PZP.....2..323G}, 
\textsc{VaST} and the online period search tool\footnote{\url{https://scan.sai.msu.ru/lk/}}. 
We have analyzed separately the data collected during the main outburst
($t_0$ to $t_0 + 8$\,d) and during the rebrightening ($t_0 + 20$\,d to $t_0 + 24$\,d).
We considered a range of trial periods (frequencies) from 0.1\,d (10\,$d^{-1}$) to 0.01\,d (100\,$d^{-1}$), 
appropriate for WZ\,Sge dwarf novae located below the period gap
\citep{1998MNRAS.298L..29K,2001ApJ...550..897H,2015PASJ...67..108K}.
The period analysis results are presented in Figures~\ref{fig:phasedlc} and \ref{fig:phasedlcbest}.

The main outburst power spectrum (Figure~\ref{fig:phasedlc}, top left panel) 
shows a weak peak corresponding to a period of 0.016209\,d ($23.34 \pm 0.03$,min). 
The analytical false alarm probability for this peak in the Lomb-Scargle periodogram is 0.01 or 0.05, 
depending on the method used to estimate the number of independent frequencies, 
following the prescriptions of \cite{2003ASPC..292..383S} and \cite{1986ApJ...302..757H}, 
respectively \citep[see discussion in][]{2008MNRAS.388.1693F,2018ApJS..236...16V}. 
We also estimate the false alarm probability of this peak using bootstrapping \citep[e.g., \S~7.4.2.3 of][]{2018ApJS..236...16V}, 
obtaining a value of 0.04 after 10,000 iterations.

This peak is not associated with a spectral window feature or any known instrumental periodicity, 
nor does it result from their combination. The interaction between a sampling feature, appearing as 
a spectral window peak at frequency $f_{\rm sampling}$, and an instrumental (or true astrophysical) signal at frequency $f_{\rm signal}$ 
can produce alias peaks at frequencies:
\begin{equation}
f_{\rm alias} = |f_{\rm signal} \pm n f_{\rm sampling}|,
\end{equation}
where $n$ is an integer \citep[e.g.,][]{2018ApJS..236...16V,2022MNRAS.513..420B}. 
If no true periodic signal is present in the data but there is slow, 
irregular variability of astrophysical or instrumental origin, 
power from these variations may leak into higher frequencies (``red-noise
leak''; e.g., \citealt{2025A&A...693A.319K}). 
Aliasing can interact with this leakage, enhancing power at unexpected locations. 
One possible scenario is that this interaction generates a series of spurious periodogram peaks, 
with lower-frequency peaks being suppressed by lightcurve detrending. In such a case, the highest-frequency peak in 
the series may remain the strongest, obscuring the fact that it is part of a series of multiple peaks.

The peak in Figure~\ref{fig:phasedlc}, top left panel, corresponds to a sine wave peak-to-mean amplitude of 0.003,mag, 
which is not distinguishable in the phased lightcurve plot (Figure~\ref{fig:phasedlc}, top right panel). 
This peak is not visible in the Lafler-Kinman string-length periodogram.

All these considerations cast doubt on the reality of the periodic signal in the main outburst lightcurve.
However, the tentative detection of this short-period signal motivated us to conduct spectroscopic observations of \source{}
revealing it as an AM\,CVn system (Section~\ref{sec:spectroscopy}).
The presence of this periodic signal during the main outburst was confirmed
with the analysis of Clear band photometry independently collect by VSNET observers
(false alarm probability estimated from bootstrap and the analytical methods: $\ll10^{-2}$).

The periodic signal becomes clearly visible during 
rebrightening (Figure~\ref{fig:phasedlc}, bottom panels) where a modulation
with a period of $P = 0.032516 \pm 0.000178$\,d ($46.82 \pm 0.26$\,min; 
twice the period found during the main outburst) and a peak-to-peak amplitude of 0.05\,mag was observed.
The larger period uncertainty compared to the one quoted above for the main outburst 
results from the quadratic relation between the period and its error 
(equation~(\ref{eq:perr})) and the shorter time baseline of INASAN-Kislovodsk $V$
band photometry during the rebrightening plateau. 
The phased lightcurve (Figure~\ref{fig:phasedlc}, bottom right) shows two
asymmetric humps per period. The peak at half that period, 
corresponding to the main outburst periodicity is also clearly seen in the
power spectrum (Figure~\ref{fig:phasedlc}, bottom left).
We find no measurable difference between the modulation period during the main
outburst (double the period corresponding to the periodogram peak in
Figure~\ref{fig:phasedlc}, top left panel) and the rebrightening.
Reassured by the absence of rapid period change, we combine all the $V$ and
Clear band data from the beginning to the end of the rebrightening to derive
the best estimate of the period (Figure~\ref{fig:phasedlcbest}):
\begin{equation}
P = 0.032546 \pm 0.000084{\rm \,d}~(46.87 \pm 0.12{\rm \,min})
\label{eq:period}
\end{equation}

We estimate the period uncertainty as 
\begin{equation}
P_{\rm err} = \Delta \phi P^2 / T,
\label{eq:perr}
\end{equation}
where $\Delta \phi = 0.5$ is the phase shift between the first and the last
points of the lightcurve that are separated by the time interval $T$
\cite[see the discussion by][of the relation between $\Delta \phi$,
``periodogram oversampling factor'' and the ``natural'' Rayleigh resolution]{2022ApJ...934..142S}. We note that while the
choice of $\Delta \phi = 0.5$ (half-a-period shift) is often too conservative
for a high signal-to-noise lightcurve, it is appropriate here given the low
amplitude of the variations in \source{}. To illustrate this, consider how
different the phased lightcurve in Figure~\ref{fig:phasedlc}, bottom right
would appear if some points were moved from phase 0.0 to phase 0.5. 
We also note that $P_{\rm err}$ determined from equation~(\ref{eq:perr}) characterize the width of an
individual periodogram peak while aliasing might inhibit one's ability to
choose the correct peak in a periodogram further limiting the period determination accuracy. 

We interpret the observed periodic modulation as positive superhumps - a type of 
photometric variations observed in some dwarf novae and related systems
\citep[e.g.,][]{2003PASP..115.1308P,2009PASJ...61S.395K,2010PASJ...62.1525K,2012PASJ...64...21K,2013PASJ...65...23K,2023MNRAS.525.1953B,2025ApJ...982..127S}. 
Positive superhumps are believed to arise from a tidal resonance in the accretion disk of the binary.
As the accretion disk grows during a dwarf nova outburst, it may extend to
the radius where the Keplerian motion in the disk is in a 3:1 resonance with
the binary orbital motion. The disk becomes eccentric due to its tidal interaction with the donor star
\citep{1988MNRAS.232...35W,1990PASJ...42..135H,1991MNRAS.249...25W}. 
The superhump period is the beat (synodic) period between the short orbital period of
the binary and the long period of disk apsidal precession -- 
the orientation of the eccentric disk relative to the donor star repeats with this period
\citep{2004MNRAS.349.1179F}. 
The tidal influence of the secondary on the distorted outer regions of the disk modulates energy dissipation in 
the disk with that period \citep{1988MNRAS.232...35W,1990MNRAS.246...29O}.
As the donor star passes the apocenter of the disk, two spiral arms are launched that travel 
inward into the disk and dissipate energy \citep{2007MNRAS.378..785S,2024A&A...689A.354J}.
Alternatively, \cite{1998ApJ...506..360S} suggest that the modulated energy production is linked 
to the accretion disk oscillating between nearly circular and highly distorted shapes over the superhump period. 
%
%
Dwarf nova outbursts that display positive superhumps are referred to as ``superoutbursts'' \citep{1995Ap&SS.226..187W}.

As the accretion disk changes its radius over the course of a superoutburst, 
the positive superhump period changes \citep{1992ApJ...401..317L}. 
The superhumps pass through 
a number of stages in their development over the curse of an outburst \citep{2009PASJ...61S.395K,2013PASJ...65..115K}.
``Stage A'' corresponds to the initial growth of the disk's eccentricity --- superhumps are just starting to appear 
(often after a short delay following the outburst onset) and have a relatively longer period (since the
disk's outer radius is maximal at this time). ``Stage B'' is the long, middle portion of the superoutburst 
where superhumps are fully developed; during this stage the superhump period often evolves slightly  
``Stage C'' refers to the tail end of the superoutburst and post-outburst phase, 
where the disk is contracting and the superhump signal weakens \citep{2012PASJ...64...21K}.
An additional phenomenon called ``early superhumps'' is observed at the very beginning of a superoutburst. 
These are low-amplitude, double-peaked modulations at (or very near) the orbital period, appearing before the ordinary (positive) superhumps develop. 
Early superhumps are believed to result from a 2:1 resonance in the disk 
-- an even more extreme tidal effect that can occur when the disk grows all the way out to where its orbital period is half the
binary's period \citep{2009PASJ...61S.395K,2023PASJ...75..619T}. 

\cite{2019PASJ...71...48I} reported the first detection of early superhumps
in AM\,CVn system NSV\,1440 (36.33\,min orbital period). 
The outburst of NSV\,1440 is similar to that of
\source{}: the first superoutburst with low-amplitude early superhumps 
(analogous to what we call ``main outburst'' in \source{}) and 
the second superoutburst (our \source{} ``rebrightening'') with ordinary superhumps
that was followed by a series of short-lived rebrightenings. 
Superhumps during rebrightenings were also observed in a hydrogen-rich WZ\,Sge dwarf nova
by \cite{2020PASJ...72...49T}.
Comparison between \source{} and NSV\,1440 suggests, that what we observed
during the main outburst might have been the early superhumps while 
the modulation during the rebrightening might be the fully-developed ``Stage B''
superhumps that in NSV\,1440 had a period very close to that of the early
superhumps. 
The lower amplitude and difference in shape between the
superhumps observed in the main outburst of \source{} and its rebrightening are consistent 
with this interpretation. The dominating half-period peak in 
the periodogram (Figure~\ref{fig:phasedlc}, top left) implies that 
the early superhumps had a shape of two identical waves per period,
markedly different from the shape of superhumps during the rebrightening 
(Figure~\ref{fig:phasedlc}, bottom panel) -- a pattern also found in
hydrogen-rich dwarf novae \citep[e.g., figures 5 and 6 of][]{2024RAA....24h5002K}.
The ``Stage A'' superhumps might have been missed due to insufficient coverage
or somehow suppressed during the dip separating the main outburst from the
rebrightening.

\cite{2016PASJ...68...55K} discuss a superoutburst of hydrogen-rich
WZ\,Sge type dwarf nova being interrupted by a dip. 
They point out that the growth time of the 3:1 resonance tidal instability (responsible for the
ordinary superhumps) is inversely proportional to the square of the binary's mass ratio
\citep{1991ApJ...381..259L}. For a sufficiently small mass ratio, 
the outburst might end by the cooling wave propagating in the accretion disk 
before the 3:1 resonance fully develops. Similarly to AM\,CVn systems NSV\,1440
\citep{2019PASJ...71...48I} and likely \source{}, the WZ\,Sge type dwarf nova ASASSN-15jd
described by \cite{2016PASJ...68...55K} showed the early superhumps before
the dip interrupting the superoutburst and ordinary superhumps after
recovering from the dip.

The prototype of the class, the AM\,CVn itself, is a non-outbursting system 
(its disk is in a permanently hot state) with an orbital period of 17\,min that 
shows three types of photometric modulation simultaneously: the positive and negative
superhumps, as well as the orbital modulation \citep{1998ApJ...493L.105H,1999PASP..111.1281S,2001MNRAS.326..621N}. 
Why the periodic signal in \source{} is a positive superhump rather than the orbital period 
or a negative superhump?
The phased lightcurve shape featuring two asymmetric humps per cycle (Figure~\ref{fig:phasedlcbest})
is very typical for superhumps. The temporal evolution of the signal (Figure~\ref{fig:phasedlc}) 
aligns with superhump behavior: the modulation emerged weakly during 
the main outburst and increased in amplitude during the rebrightening. 
An orbital signal, by contrast, should persist independently of outburst
state. The same is true for negative superhumps -- another kind of
variation attributed to a tilted accretion disk experiencing a nodal precession in a retrograde
direction \citep{2009MNRAS.394.1897M,2010A&A...514A..30T}. 
The tilted geometry changes the way in which the accretion stream from the secondary star interacts with the disk. 
Instead of always striking the outer edge in a nearly constant geometry, the stream may alternatively pass over or under the
disk, depositing kinetic energy at different annuli in a time-variable fashion.
As a consequence, the observer sees brightness modulations with a period that is modestly shorter than the orbital period.
Negative superhumps are typically present in quiescence and persist over a dwarf nova outburst
cycle \citep[e.g.,][]{2010OAP....23...98S,2013PASJ...65...50O}.
In summary, the periodic signal observed in \source{} is interpreted as positive superhumps rather than 
an orbital modulation or negative superhumps based on analogy with other well-studied AM\,CVn
systems \citep{1989MNRAS.236..319O,harrop-allin1996superhumps} where the orbital period is known from
spectroscopy \citep{2018MNRAS.476.1663G} and eclipses \citep{2015gacv.workE..25B}.

\subsection{Recognizing \source{} as AM\,CVn}
\label{sec:class}

After the discovery of the outburst (Sec.~\ref{sec:nmw}) we initially
assumed that \source{} is an ordinary (hydrogen-rich) dwarf nova of WZ\,Sge subtype. 
WZ\,Sge stars are characterized by rare (once in a decade or more) 
outbursts with a median amplitude of 7.7\,mag \citep[see][for a review]{2015PASJ...67..108K}. 
Such systems are often found in optical transient surveys, especially the ones aimed at finding classical 
novae \citep[e.g.,][]{2019CoSka..49..204P,2020PASJ...72...49T,2021AJ....161...15S,2024RAA....24h5002K,2025arXiv250207447K}. 
Before a definitive spectroscopic classification is available, 
WZ\,Sge systems can be distinguished from classical novae by their blue 
color, lower absolute magnitude (if Gaia parallax is available) and 
the appearance of superhumps in their lightcurves a few days after 
the eruption.




While the color and absolute magnitude of \source{} appeared consistent 
with expectations for a WZ\,Sge-type dwarf nova, the absence of evident superhumps 
with a typical dwarf nova period and the tentative detection of a shorter
23\,min periodicity (Section~\ref{sec:period}) suggested that \source{} might be something else. 
The presence of the short-period signal, the outburst amplitude and duration being shorter 
than those typically found in WZ\,Sge systems, and the overall similarity of \source{} to 
the recently studied AM\,CVn system ASASSN-21br \citep{2024MNRAS.532.4205P} motivated us to conduct spectroscopic observations.

\subsection{Comparison with other AM\,CVn binaries}
\label{sec:compare}

Peaking at $V_{\rm max} = 12.45$ (Table~\ref{tab:mags}), \source{} is among the five brightest AM\,CVn
outbursts ever observed along with ASASSN-14mv, ASASSN-14ei, SDSS\,J141118.31$+$481257.6
and NSV\,1440 \citep{2018A&A...620A.141R}.
The peak absolute magnitude $M_V = 3.4^{+0.7}_{-0.9}$ is about 2\,mag brighter than those displayed in fig.~3 of \cite{2018A&A...620A.141R}.
The unusually bright absolute magnitude during the outburst (while the
quiescence magnitude is fairly typical) may be an indication that the
accretion disk in \source{} is viewed almost face-on which maximizes its 
apparent brightness \citep{1980AcA....30..127P,2011MNRAS.411.2695P}.
%
According to the equation~(12) of \cite{2002A&A...383..574O}, the amplitude of early superhumps is determined by the inclination of the system,
with lower inclination resulting in smaller amplitude. Thus, the face-on interpretation is also consistent with the small
amplitude of early superhumps observed in \source{} (\S~\ref{sec:period}).
The intrinsic colors of \source{} at minimum are 
$(g-r)_0 = -0.24$,
$(BP-RP)_0 = -0.14$ (Table~\ref{tab:mags}).
The quiescent absolute magnitude and colors of \source{} are typical for AM\,CVn
binaries \citep[see fig.~7 of][]{2021MNRAS.505..215R}. 

\begin{figure}
\centering
\includegraphics[width=1.0\columnwidth,clip=true,trim=0.0cm 0.0cm 0.0cm 0.0cm,angle=0]{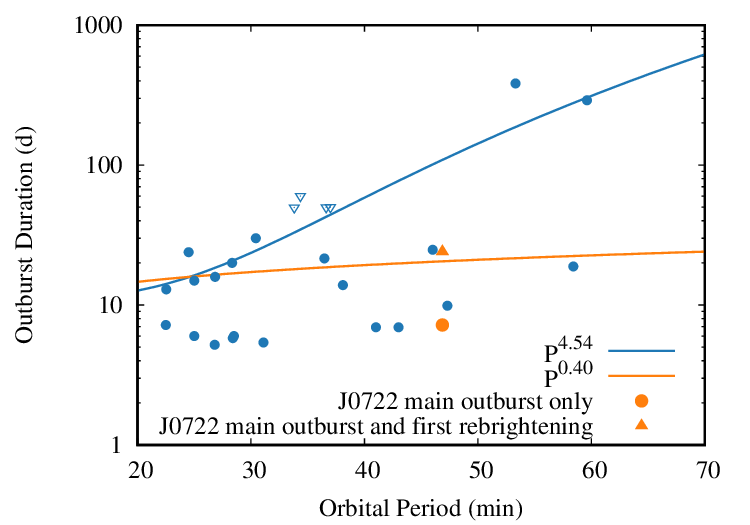}
\caption{Outburst duration as a function of the orbital period for AM\,CVn
stars from \cite{2019AJ....157..130C,2021MNRAS.508.3275P,2022ApJ...926...10R,2024MNRAS.532.4205P}.
For systems where \cite{2019AJ....157..130C} and \cite{2021MNRAS.508.3275P}
give different outburst durations (KL\,Dra, CP\,Eri) both estimates are
plotted as they are presumably based on different outbursts. 
Open triangles represent upper limits on the outburst duration.
The top blue curve represents the empirical relation of \cite{2015MNRAS.446..391L}.
The bottom orange curve is the accretion disk limit cycle model of \cite{2015ApJ...803...19C,2019AJ....157..130C}.}
\label{fig:dur_porb}
\end{figure}

The characteristic time over which matter moves radially through the
accretion disk, the viscous timescale, 
may directly correspond to the duration of a superoutburst according 
to \cite{2019AJ....157..130C,2021A&A...650A.114H}, 
while other authors suggested this may be the case for normal outbursts 
\citep{2012A&A...544A..13K,2021MNRAS.508.3275P}.
Following \cite{2015MNRAS.446..391L}, \cite{2019AJ....157..130C} and \citep{2021MNRAS.508.3275P}, 
we consider the relation between the outburst duration and orbital period, 
putting \source{} on the plot (Figure~\ref{fig:dur_porb}) using its
superhump period as the proxy for the orbital period. 
We indicate two locations of \source{} on the plot: one if only duration of 
the main outburst is considered and the other if the time between the start of
the main outburst and the decline from the first rebrightening is taken as
the outburst duration. A well-sampled long-duration lightcurve
is needed to make this distinction \citep{2021MNRAS.508.3275P}. A confusion
between the main outburst and a rebrightening may increase the scatter of 
outburst duration estimates in Figure~\ref{fig:dur_porb}. 
At least two other AM\,CVn systems, NSV\,1440 \citep{2019PASJ...71...48I}
and SDSS\,J141118.31$+$481257.6 \citep{2019MNRAS.483L...6R,2019AJ....157..130C}
show the first rebrightening characterized by a long-lived plateau 
(resembling a superoutburst of a dwarf nova) followed by a series of
short-lived rebrightenings (resembling ordinary dwarf nova outbursts). 
According to \cite{2019AJ....157..130C}, it is unclear if the first
superoutburst-like rebrightening should be included in the outburst duration 
for the purposes of $\tau_{\rm visc}$ estimation (Figure~\ref{fig:dur_porb}).
Shorter-period AM\,CVn systems often show a lightcurve feature described as ``dip'' 
in their main outburst lightcurve \citep{2012MNRAS.425.1486R,2021MNRAS.502.4953D} 
which may be a less dramatic version of the fading that separates the main outburst from 
the first rebrightening in \source{}, SDSS\,J141118.31$+$481257.6 and NSV\,1440. 

%

No X-ray source is associated with \source{}. 
The Upper Limit Server\footnote{\url{http://xmmuls.esac.esa.int/hiligt/}} \citep{2011ASPC..442..567S} 
provides a $2\sigma$ 0.2--2\,keV ROSAT/PSPC upper limit of 0.026\,cts/s
from the survey \citep{1999A&A...349..389V,2000IAUC.7432....3V,2016A&A...588A.103B} 
observations performed on 1990-09-15.
The V band extinction, $A_V$, listed in Table~\ref{tab:mags} corresponds to 
the equivalent hydrogen absorbing column $N_\mathrm{H} = 6.63 \times 10^{20}\,{\rm cm}^{-2}$ 
according to the \cite{2009MNRAS.400.2050G} calibration. For this
$N_\mathrm{H}$ value and a fiducial powerlaw spectrum with a photon index $\Gamma = 2$ 
the ROSAT/PSPC count rate upper limit corresponds to a 0.2--2\,keV flux
upper limit of $5\times10^{-13}$\,erg\,cm$^{-2}$\,s$^{-1}$. 
The corresponding luminosity upper limit $L_X < 2\times10^{31}$\,erg\,s$^{-1}$ is
one order of magnitude above the average X-ray luminosity of AM\,CVn systems
\citep{2025PASP..137a4201R}, so the X-ray non-detection is not in tension
with the AM\,CVn classification.
However, a number of X-ray bright AM\,CVn systems have X-ray luminosities
above the \source{} upper limit \citep{2006A&A...457..623R,2023JAVSO..51..227B}.
The X-ray flux of AM\,CVn systems may vary with time by a factor of a few 
\citep{2012MNRAS.425.1486R,2019MNRAS.483L...6R}.

\section{Conclusions}
\label{sec:conclusion}

We report the discovery and spectroscopic confirmation of TCP\,J07222683$+$6220548, an AM\,CVn system that
displayed a single $\Delta V = 7.6$\,mag outburst followed by multiple rebrightenings in January-March 2025. 
The overall shape and duration of the outburst are similar to those found in long-period AM\,CVn stars. 
A periodic modulation in the lightcurve, equation~(\ref{eq:period}),  
that is barely detectable during the main outburst and grows in peak-to-peak amplitude 
to 0.05\,mag during the first rebrightening is interpreted as positive superhumps (with two waves per period).
No change in superhump period is observed over the course of their development.
As few dwarf novae benefit from a detailed photometric and 
spectroscopic follow-up similar to the one performed for \source{}, 
it is possible that some outbursting AM\,CVn systems remain unrecognized
among dwarf novae candidates identified by surveys.

\begin{acknowledgements}

We thank the anonymous referee for the helpful comments. 
KS is grateful to Dr.~Alexandra Zubareva for an illuminating discussion on the physical origin of superhumps.
The observations were performed at two telescopes Astrosib RC-500 of shared research facility ``Terskol observatory'' 
of Institute of Astronomy of the Russian Academy of Sciences. 
We acknowledge with thanks the variable star observations from 
the Variable Star Observers League in Japan (VSOLJ) database 
and the AAVSO International Database contributed by observers worldwide and used in this research. Scientific equipment used in this study was bought partially through the M. V. Lomonosov Moscow State University Program of Development. We are grateful to the staff of CMO SAI for their assistance in organizing alert observations.
A.~Tarasenkov acknowledges the support of the Foundation for the Development of Theoretical Physics and Mathematics BASIS (project 24-2-1-6-1).
This research has made use of the International Variable Star Index (VSX) database, operated at AAVSO, Cambridge, Massachusetts, USA, 
the Aladin sky atlas \citep{1999ASPC..172..229B} and 
the VizieR catalogue access tool \citep{vizier2000} developed at CDS, Strasbourg, France. 
Lasair is supported by the UKRI Science and Technology Facilities Council and is a collaboration between 
the University of Edinburgh (grant ST/N002512/1) and Queen's University Belfast (grant ST/N002520/1) within the LSST:UK Science Consortium.
Based on observations obtained with the Samuel Oschin Telescope 48-inch and the 60-inch Telescope at the Palomar
Observatory as part of the Zwicky Transient Facility project. ZTF is supported by the National Science Foundation under Grants
No. AST-1440341 and AST-2034437 and a collaboration including current partners Caltech, IPAC, the Oskar Klein Center at
Stockholm University, the University of Maryland, University of California, Berkeley, the University of Wisconsin at Milwaukee,
University of Warwick, Ruhr University, Cornell University, Northwestern University and Drexel University. Operations are
conducted by COO, IPAC, and UW.
This work has made use of data from the Asteroid Terrestrial-impact Last Alert System (ATLAS) project. The Asteroid Terrestrial-impact Last Alert System (ATLAS) project is primarily funded to search for near earth asteroids through NASA grants NN12AR55G, 80NSSC18K0284, and 80NSSC18K1575; byproducts of the NEO search include images and catalogs from the survey area. This work was partially funded by Kepler/K2 grant J1944/80NSSC19K0112 and HST GO-15889, and STFC grants ST/T000198/1 and ST/S006109/1. The ATLAS science products have been made possible through the contributions of the University of Hawaii Institute for Astronomy, 
the Queen's University Belfast, the Space Telescope Science Institute, the South African Astronomical Observatory, and The Millennium Institute of Astrophysics (MAS), Chile.
This research has made use of the Astrophysics Data System, funded by NASA under Cooperative Agreement 80NSSC21M00561.

\end{acknowledgements}

\bibliographystyle{raa}
\bibliography{j0722}

\label{lastpage}

\end{document}